\documentclass[showpacs,a4paper,amsmath,amssymb]{revtex4}
\usepackage{graphicx}% Include figure files
\usepackage{dcolumn}% Align table columns on decimal point
\usepackage{bm}% bold math
\begin{document}
%\bibliographystyle{unsrt}
%-----------------------------------------------------------------------
%-                            My Title                                 -
%-----------------------------------------------------------------------
\title{\bf Dust and Radiation Quantum Perfect Fluid Cosmology : Selection of Time Variable}
\author{I-Chin Wang}
\email{icwang@hepth.phys.ncku.edu.tw} \affiliation{Department of
Physics, National Cheng Kung University\\ Tainan, Taiwan 701}
\date{\today}
%-----------------------------------------------------------------------
\begin{abstract}
We studied the expectation value of the scale factor in radiation
and dust quantum perfect fluid cosmology. We used Schutz
variational formalism to describe perfect fluid and selected the
conjugate coordinate of perfect fluid be dynamical variable. After
quantization and solving the Wheeler-DeWitt equation can obtain
the exact solution. By superposition of the exact solution, we
obtained one wave packets and used it to compute the expectation
value of the scale factor. We found that if one select different
dynamical variable be the time variable in each of these two
systems, the expectation value of the scale factor of these two
systems can fit in with the prediction of General Relativity.
Therefore we thought that the selection of reference time can be
different for different quantum perfect fluid systems.

\end{abstract}
\maketitle
\section{Introduction}\label{Sec.1}
It is known that time disappear in the canonical quantization
based on Arnowitt-Deser-Misner(ADM) decomposition. The notion of
time can be recovered in some case of quantum cosmology, for
example when gravity were coupled to a perfect
fluid~\cite{Rubakov:77,Alvarenga:02,Batista:02}. These kind of
systems were often dealt with in the following way
(see~\cite{Alvarenga:02,Batista:02}): First they used Schutz's
formalism for the description of perfect fluid~\cite{Schutz:70},
second they selected the dynamical variable of perfect fluid be
reference time. Then used canonical quantization to obtain the
Wheeler-Dewitt equation in minisuperspace, which is a
Schr\"{o}dinger like equation~\cite{Rubakov:77}. After solving the
equation and choosing a weighting, they can obtain one wave
packets. The obtained wave packets can be used to compute the
expectation value of the scale factor. By using the wave packets
they obtained the time-dependent behavior of the scale factor. If
the selected time variable(dynamical variable of perfect fluid)
can let the expectation value of the scale factor fit in with the
classical prediction(prediction of General Relativity) when time
is large, the selected time variable can be viewed as a good time
variable for the system.

The above method can reconstruct a time variable and let the
obtained expectation value fit in with GR's prediction, but there
are still some questions. First, although the solution for both
the radiation-dominated quantum PF(perfect fluid) system and the
dust-dominated quantum PF system can be obtained, the good
weighting function is hard to find. In order to avoid this
problem, Lemos~\cite{Lemos:96} proposed another procedure. He
chosen a special initial wave packets satisfied the dynamical
equation of the system. Then he used the wave packets and the
propagator to compute the dynamical evolution of the scale factor,
 the behavior of the scale factor did fit in with classical
prediction.

Although this problem had been solved by Lemos but there is
another way to discuss this problem. In this article we try to
discuss these two systems in the other way--by choosing a
weighting and a suitable time variable. We obtained the exact
solution of radiation-dominated and dust-dominated quantum PF
cosmology. We found that for the radiation-dominated system, if
the square of the perfect fluid dynamical variable can be treated
as time variable, there exist a suitable weighting function that
can reconstruct a wave packets which make the expectation value of
the scale factor fit in with classical prediction. While for the
dust-dominated quantum PF cosmology, there exist another suitable
weighting function which make the perfect fluid dynamical variable
be a good time variable. In other words, different weighting
function corresponds to different choice of time variable.

This paper is organized as follows. In the next section, the
quantum perfect fluid cosmological model in minisuperspace was
established. By using Schutz's formalism, the Wheeler-DeWitt
equation for arbitrary barotropic equation of state $p=\alpha\rho$
is written down. In section 3, we solved the exact solution for
radiation and dust universe in flat case. After choosing a
weighting function, we obtain the wave packets. Then we use the
wave packets to compute the expectation value of the scale factor
of universe.  We found that the time variable and the weighting
function can be chosen differently for these two different system.
For dust-dominated system, the perfect fluid dynamical variable is
a good time variable. While for radiation-dominated system, the
square of the perfect fluid dynamical variable is a suitable time
variable. Section 4 is discussion.

\section{The Quantum Model}\label{Sec.2}
The action for a perfect fluid coupled to gravity in Schutz's
formalism can be described as
\begin{equation}
S=\int_M d^{4} x \sqrt{-g} p-\int_M d^{4} x\sqrt{-g}
R-\int_{\partial M} d^{3} x \sqrt{h} K
\end{equation}
where p is the pressure, K is the trace of the extrinsic curvature
and h is the induced metric in the boundary of the manifold M. The
pressure p is linked to the energy density by the equation of
state $p=\alpha\rho$. The four velocity of the fluid is expressed
in terms of the five potentials $\epsilon$, $\phi$, $\beta$,
$\theta$ and $S$:
\begin{equation}
u_\nu=\frac{1}{\mu}(\epsilon_,\nu +\phi\beta_,\nu +\theta S_,\nu)
\end{equation}
where $\mu$ is the specific enthalpy. The four velocity is
subjected to the condition $u^{\nu} n_{\nu} =1$. We consider the
Friedmann-Lemaitre-Robertson-Walker metric
\begin{equation}
ds^2=-N^2 dt^2 +a^2(t)\sigma_{ij} dx^i dx^j
\end{equation}
the super-hamiltonian can be constructed
as~\cite{Rubakov:77,Alvarenga:02,Batista:02}
\begin{equation}
H=-\frac{\Pi^{2}_a}{24a}-6ka+\frac{e^{S}
p^{1+\alpha}_\epsilon}{a^{3\alpha}}
\end{equation}
where $\alpha$ was mentioned above as the equation of state
($p=\alpha\rho$). Performing the canonical
transformations~\cite{Rubakov:77,Alvarenga:02,Batista:02}
\begin{equation}
T_2=-p_se^{-s}p^{-(1+\alpha)}_{\epsilon}\hspace{5mm},\hspace{5mm}
\Pi_{T_2}=e^sp^{1+\alpha}_{\epsilon}
\end{equation}
where $T_2$ is the dynamical variable for perfect fluid,
$\Pi_{T_2}$ is the conjugate momentum of $T_2$. The
super-hamiltonian can be written as
\begin{equation}
H=-\frac{\Pi^{2}_a}{24a}-6ka+\frac{\Pi_T}{a^{3\alpha}}
\end{equation}
note that $\Pi_{T_2}$ is also the momentum associated with the
matter variable, $k$ represents the curvature of the 3-space with
the values 0, 1, -1 for a flat, closed and open Universe.
Introducing the quantization conditions to the hamiltonian, we
obtain the Wheeler-DeWitt equation in the minisuperspace:
\begin{equation}
\frac{\partial^2 \Psi}{\partial a^2}-144ka^2-i24a^{1-3\alpha}
\frac{\partial\Psi}{\partial {T_2}} =0
\end{equation}
the equation (6) is equivalent to the Schr\"{o}dinger equation of
ordinary quantum mechanics. Thus, all the formalism of quantum
mechanics can be applied to this problem. Applying the separation
of variables method, writting
\begin{equation}
\Psi(a,T)=e^{iET} \psi(a)
\end{equation}
we obtain
\begin{equation}
\varphi''+(-144ka^2 +24Ea^{1-3\alpha})\psi=0
\end{equation}
where the primes mean derivative with respect to a. So far, we
obtain the equation for 3-space. After defining the curvature of
the 3-space and the matter which is filled in this Universe, we
will obtain the dynamical equation for the individual Universe.
 \section{Behavior of the scale factor with the selection of time variable}
 \label{Sec.3}
We consider the flat case (k=0). Thus equation (9) reduces to
\begin{equation}
\psi''+24Ea^{1-3\alpha} \psi=0
\end{equation}

Although the general solution of equation (10) has been
obtained~\cite{Batista:02,Alvarenga:02}, but this general solution
can not become radiation-dominated solution when
$\alpha=\frac{1}{3}$. We first consider radiation-dominate case.

\subsection{Radiation
($\alpha=1/3$)}\label{Sec.3.1}
 For $\alpha=\frac{1}{3}$ ,
equation (10) reduced to
\begin{equation}
\psi''(a)+24E\psi(a)=0
\end{equation}
the solution is
\begin{equation}
\psi(a)=D_1sin(\sqrt{E}a)+D_2cos(\sqrt{E}a)
\end{equation}
note that the solution is not the form of Bessel function. We
choose $D_1$=0 to construct wave packets, and our boundary
condition~\cite{Lemos:96} for wave packets is
\begin{equation}
\left.\frac{\partial\Psi(a,T_2)}{\partial a}\right|_{a=0}=0
\end{equation}
put $\psi(a)=D_2cos(\sqrt{E}a)$ into equation (8). To obtain a
wave packets, we choose a weighting function $W(E)$
\begin{equation}
\Psi(a,T_2)=\int_0^{-\infty}e^{iET_2}cos(\sqrt{E}a)W(E)dE
\end{equation}
where $W(E)$ is the weighting. Let $E=y^2$ and take
$W(y)=y^{-1}e^{-\lambda y^2}$ , then put these into equation (12)
\begin{equation}
\Psi(a,T_2)=\int_0^{-\infty}2e^{-(\lambda-iT_2)y^2}cos(ya)dy
\end{equation}
note that Re$\lambda>0$, use formula ~\cite{formula:3.896-4} to
get the wave packets
\begin{equation}
\Psi(a,T_2)=-\sqrt{\frac{\pi}{(\lambda-iT_2)}} exp
\{-\frac{a^2}{4(\lambda-iT_2)}\}
\end{equation}
the inner product must take the form in order for the Hamiltonian
operator $\hat{H}$ to be self-adjoint
\begin{equation}
(\Phi,\Psi)=\int_0^{\infty}a^{1-3\alpha}\Phi^* \Psi da
\end{equation}
the expectation value of the scale factor is
\begin{equation}
<a>(T_2)=\frac{\int_0^{\infty} \Psi^{*}(a,T_2)a\Psi(a,T_2)
da}{\int_0^{\infty} \Psi^{*}(a,T_2)\Psi(a,T_2)da}
\end{equation}
leading to
\begin{equation}
<a>(T_2)=\frac{\sqrt{\lambda^2+T_2^2}}{\sqrt{2\lambda\pi}}\propto{T_2}
\end{equation}
Now we choose $T_2=\sqrt{T}$ , that $T$ is our time variable. When
$T\rightarrow\infty$
 the expectation value of $<a>$ becomes
\begin{equation}
<a>(T)\propto{T^{\frac{1}{2}}}
\end{equation}
the behavior of the scale factor for radiation is proportional to
half order of $T$, just fit in with classical prediction. In this
case with this weighting function $W(y)$, we see that $T_2^2$ is
the suitable time variable. Next we discuss dust-dominated system.

\subsection{Dust
($\alpha=0$)}\label{Sec.3.1}  The solution of equation (9) for
$\alpha=\frac{1}{3}$ had been
obtained~\cite{Batista:02,Alvarenga:02,Lemos:96}. We review this
dust-dominated system to see which dynamical variable could be
time variable.

For $\alpha=0$ , equation (9) reduce to
\begin{equation}
\psi''(a)+24Ea\psi(a)=0
\end{equation}
the solution is the form of Bessel function
\begin{equation}
\psi(a)=
\sqrt{z}\{c_1J_{\frac{1}{3}}(\frac{2}{3}z^{\frac{3}{2}})+c_2J_{\frac{-1}{3}}(\frac{2}{3}z^{\frac{3}{2}})\}
\end{equation}
where $z=-E^{\frac{1}{3}}a$
\begin{equation}
\Psi_E=e^{iET_2}
\sqrt{z}\{c_1J_{\frac{1}{3}}(\frac{2}{3}z^{\frac{3}{2}})+c_2J_{\frac{-1}{3}}(\frac{2}{3}z^{\frac{3}{2}})\}
\end{equation}
we choose $c_2=0$ and a weighting function $W'(E)$ to construct
the wave packet, the boundary condition for the wave packets is
the same as e.q.(13)
\begin{equation}
\Psi(a,T_2)=\int_0^{-\infty} W'(E)\Psi_E(a,T_2) dE
\end{equation}
let $E=y^2$ , $W'(y)=e^{-\lambda y^{2}}$(note Re$\lambda>0$) and
use formula ~\cite{formula:6.631-4} , the wave packets is
\begin{equation}
\Psi(a,T_2)=\frac{c_4}{\{2(\lambda-iT_2)\}^{\frac{4}{3}}}
exp\{\frac{-c_3^2}{4(\lambda-iT_2)}\}
\end{equation}
where $c_3=-\frac{2}{3}a^{\frac{3}{2}} $ , $c_4=\frac{2}{3} c_1
(1-i\sqrt{3})a$ , the expectation value of the scale factor
(for$\alpha=0$) is
\begin{equation}
<a>(T_2)=\frac{\int_0^{\infty} a\Psi^{*}(a,T_2)a\Psi(a,T_2)
da}{\int_0^{\infty}
a\Psi^{*}(a,T_2)\Psi(a,T_2)da}\propto(\lambda^2+\frac{T_2^2}{\lambda})^{\frac{1}{3}}
\end{equation}
when $T_2\rightarrow\infty$ :
\begin{equation}
<a>(T_2)\propto{T_2^{\frac{2}{3}}}
\end{equation}
Thus for dust-dominated quantum cosmology $(\alpha=0)$ with the
weighting function $W'(y)$, $T_2$ ( the dynamical variable for
dust perfect fluid ) is a suitable time variable because it makes
the behavior of scale factor fit in with classical prediction.

Thus we can see that the suitable time variable depends on the
weighting function we chosen.
\section{ Discussion }\label{Sec.4}

If we think that time disappear in the canonical quantization
based on Arnowitt-Deser-Misner(ADM) decomposition can be solved by
choosing a dynamical variable as the time variable, then we can
find that the choice is not unique. In section 3, we can see that
a suitable time variable depends on the weighting function one
used.

Note that in Lemos's paper, the situation is different. Lemos
chosen a suitable wave packets to satisfy his time variable (the
dynamical variable for perfect fluid). In other words, he first
decided the time variable of the system then search a good wave
packets to make the behavior of the scale factor fit in with
classical prediction.

Therefore in the canonical quantization based on ADM
decomposition, the choice of time variable is not unique but is
relative. The wave packets we used did affect the time variable we
chose. In this point of view, time is not an absolute coordinate,
it is a relative parameter.

As for the parameter $t$ appeared in the equation (3), we consider
that $t$ is an introduced parameter which helps us to think of a
physical system. We think that this is the case Barbour et al
proposed~\cite{Barbour:02}. In the article there are two
sentence~\cite{Barbour:02.1}:
\begin{quotation}\it
The theory of time and clocks must be derived from the dynamics of
the universe. The fact is that speed is not obtained by dividing
an infinitesimal displacement by an infinitesimal time but by
another displacement.
\end{quotation}
Thus we think that time variable can be chosen and in the
canonical quantization based on ADM decomposition, it depends on
the wave packets we used.

%\bibliography{myreference}

\end{document}